\documentclass{article}

\usepackage{PRIMEarxiv}

\usepackage[utf8]{inputenc} 
\usepackage[T1]{fontenc}    
\usepackage{url}            
\usepackage{booktabs}       
\usepackage{amsfonts}       
\usepackage{nicefrac}       
\usepackage{microtype}      
\usepackage{lipsum}
\usepackage{fancyhdr}       
\usepackage{graphicx}       
\graphicspath{{media/}}     

\usepackage{amsmath}
\usepackage{float}
\usepackage[hidelinks]{hyperref}

\title{Citation Recommendation using Deep Canonical Correlation Analysis
}

\author{
  Conor J. McNamara \\
  School of Computer Science \\
  University of Galway \\
  County Galway, Ireland\\
  \texttt{c.mcnamara33@universityofgalway.ie} \\
   \And
  Effirul I. Ramlan \\
  School of Computer Science \\
  University of Galway \\
  County Galway, Ireland\\
  \texttt{effirul.ramlan@universityofgalway.ie} \\
}

\begin{document}
\maketitle

\begin{abstract}
Recent advances in citation recommendation have improved accuracy by leveraging multi-view representation learning to integrate the various modalities present in scholarly documents. However, effectively combining multiple data views requires fusion techniques that can capture complementary information while preserving the unique characteristics of each modality. We propose a novel citation recommendation algorithm that improves upon linear Canonical Correlation Analysis (CCA) methods by applying Deep CCA (DCCA), a neural network extension capable of capturing complex, non-linear relationships between distributed textual and graph-based representations of scientific articles. Experiments on the large-scale DBLP (Digital Bibliography \& Library Project) citation network dataset demonstrate that our approach outperforms state-of-the-art CCA-based methods, achieving relative improvements of over 11\% in Mean Average Precision@10, 5\% in Precision@10, and 7\% in Recall@10. These gains reflect more relevant citation recommendations and enhanced ranking quality, suggesting that DCCA’s non-linear transformations yield more expressive latent representations than CCA’s linear projections.
\end{abstract}


\section{Introduction}
With the growing volume of scientific literature published each year, researchers face increasing challenges in identifying and filtering relevant articles within their fields \cite{Beel2015Review}. Traditionally, researchers have relied on manual methods for literature discovery, such as keyword searches on scholarly platforms like Google Scholar\footnote{\url{https://scholar.google.com}} or tracing citations between papers. While these approaches can be effective, they are often time-consuming and limited in scope, requiring researchers to evaluate articles individually. Moreover, keyword searches may fail to capture the nuances of researchers' information needs \cite{ElAriniBeyondKeywordSearch}. Recently, applications utilizing large language models (LLMs), such as Perplexity\footnote{\url{https://www.perplexity.ai}}, have gained attention as tools for literature discovery, offering rapid access to potentially relevant articles. However, their use introduces trade-offs, including concerns about accuracy, coverage, and interpretability \cite{Noorden2023AI}.
 
Citation recommendation systems address these limitations by generating ranked lists of relevant publications to cite, tailored to a given query paper or text segment. These systems help streamline the literature review process, allowing researchers to identify relevant works more efficiently. Current research in this area can be categorized into four main information filtering approaches: collaborative filtering, content-based, graph-based, and hybrid methods \cite{Ali2021Review}. Collaborative filtering models user-document interactions, but often faces cold-start and data sparsity issues. Content-based approaches analyze the textual features of papers, but typically overlook the relational context between entities such as papers, authors, and venues. Graph-based methods leverage network structures to capture these relationships, though they often ignore rich content information. Hybrid approaches seek to overcome these shortcomings by integrating multiple information filtering strategies and leveraging complementary strengths to improve recommendation performance. Consequently, such methods have become increasingly popular in recent studies.

Among hybrid approaches, multi-view representation learning has gained considerable attention for its ability to integrate heterogeneous data sources. Canonical Correlation Analysis (CCA) \cite{Hardoon2004CCA} has been used to learn linear transformations that project text and node embeddings of scientific articles into a shared subspace where their correlation is maximized. Linearly combining these projections to form fused representations has demonstrated improved performance compared to single-view methods \cite{Gupta2017TextAndGraph, Chanana2019TextAndGraph}. However, the linear nature of CCA limits its ability to capture complex, non-linear relationships between data views \cite{Andrew2013DCCA}, potentially leading to suboptimal latent representations. To address this, we propose using Deep Canonical Correlation Analysis (DCCA) \cite{Andrew2013DCCA}, an extension of CCA that leverages neural networks to model non-linear relationships between modalities. We hypothesize that DCCA can learn more expressive latent representations and enhance citation recommendation performance. DCCA has demonstrated success in various domains, such as user recommender systems \cite{Jirachanchaisiri2022DCCA}, image processing \cite{Yan2015DCCA}, and speech recognition \cite{Isobe2021DCCA}, but it has not yet been applied to citation recommendation tasks.

The main contributions of this work are as follows:
{
\renewcommand{\labelitemi}{--}
\begin{itemize}
    \item We introduce the use of DCCA for citation recommendation, a novel application not explored in prior work.
    
    \item We propose a multimodal algorithm that integrates text and node embeddings of scientific articles into a shared latent space using DCCA.

    \item We evaluate our approach on the large-scale DBLP citation network dataset and demonstrate that it outperforms state-of-the-art CCA-based citation recommendation methods in terms of precision, recall, and mean average precision.
\end{itemize}
}

The remainder of this paper is organized as follows: Section \ref{sec:literature-review} provides a literature review of the field. Section \ref{sec:proposed-approach} details our proposed methodology. Section \ref{sec:experimental-setup} outlines the experimental setup. Section \ref{sec:results} presents and discusses the experimental results. Finally, Section \ref{sec:conclusion} concludes the paper and suggests directions for future research.

\section{Literature Review}
\label{sec:literature-review}
This section provides a comprehensive review of the field of citation recommendation, highlighting its key tasks, approaches, and evaluation techniques.

\subsection{Citation Recommendation}
Citations are an essential aspect of academic writing because they establish credibility, support claims, and acknowledge prior contributions. Although the terms 'citation' and 'reference' are often used interchangeably, a citation refers to the mention of a source of information within a text, while a reference provides the full details of that source, usually in a reference list \cite{Farber2020Review}.

Citation recommendation is an extension of the paper recommendation task. While paper recommendation aims to recommend publications that may interest a user and are worthwhile to read \cite{Beel2015Review}, citation recommendation specifically focuses on assisting users in substantiating their work by recommending publications to use as citations \cite{Farber2020Review}. The citation recommendation problem can be viewed as learning a scoring function $f(p,q)$ that evaluates the relevance of a target paper {$p \in P$} to a query paper {$q \in Q$}, where {$Q$} is the set of query papers and {$P$} is the set of target papers. The function assigns a score that indicates how relevant $p$ is as a citation for $q$. Citation recommendation systems are particularly useful during the writing phase of a paper or when conducting a comprehensive literature review. They can also assist reviewers and editors in assessing the relevance and completeness of references cited in a manuscript.

\subsection{Citation Recommendation Tasks}
\label{sec:tasks}
\begin{figure}[h]
    \centering
    \includegraphics[width=1\textwidth]{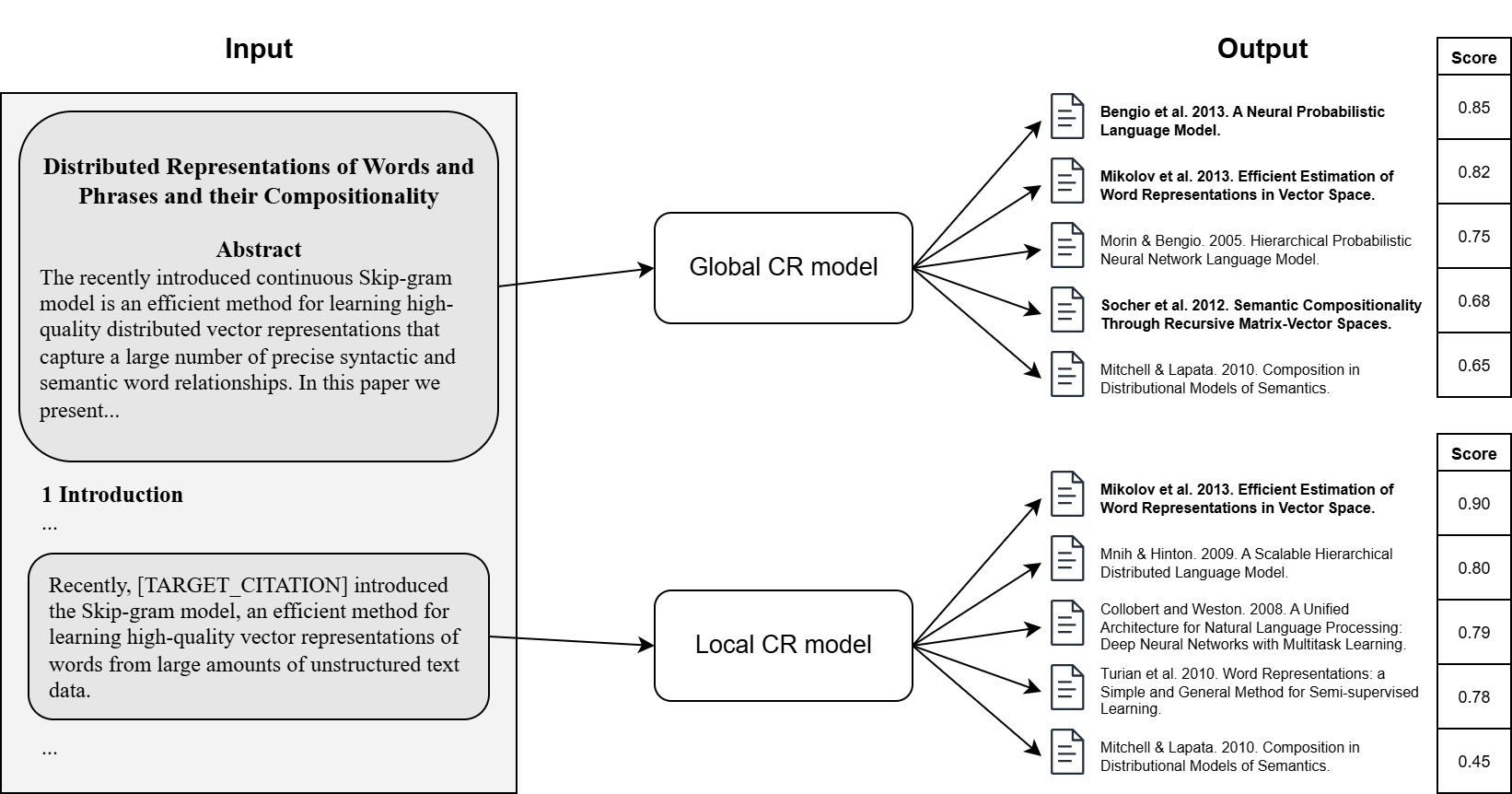}
    \caption{Comparison of the difference between global and local citation recommendation (CR) systems. The top section shows a typical input to a global CR model: the title and abstract of the citing article. The bottom section depicts a typical input to a local CR model: a citation context where the the citation is masked with a placeholder ("TARGET\_CITATION"). Both systems output a ranked list of articles ordered by their relevance scores. Titles shown in bold indicate correct recommendations. Reproduced from \cite{Medic2021Review} with permission, licensed under CC BY-ND 4.0.}
    \label{fig:global_vs_local}
\end{figure}

Citation recommendation tasks are generally categorized as either global or local, depending on the range of context used \cite{He2010ContextAware}. Our proposed method focuses on global citation recommendation. Global citation recommendation involves recommending citations for an entire manuscript, typically using its title, abstract, or full text as input, aiming to provide a broad set of relevant citations. In contrast, local citation recommendation, also referred to as context-aware citation recommendation, focuses on recommending citations for specific citation markers (e.g., "[1]") within manuscripts using the surrounding text as context, resulting in more focused recommendations. Figure \ref{fig:global_vs_local} illustrates the key differences between these two approaches. Notably, global systems often have multiple correct recommendations, while local systems typically have only one, since they target individual citation contexts.

\subsection{Citation Recommendation Approaches}
\subsubsection{Collaborative Filtering-Based Citation Recommendation}
Collaborative filtering is a classic recommendation technique that recommends items by analyzing the preferences or behaviors of users with similar interests. It can be broadly categorized into memory-based and model-based approaches \cite{Aggarwal2016Recommender}.

Memory-based methods use historical user-item interactions to identify similarities between users or items. In user-based collaborative filtering, recommendations are generated by finding users with similar preferences and recommending items to the target user that they have not interacted with. In item-based collaborative filtering, recommendations are made by identifying items similar to those that a user has previously engaged with. Common similarity measures to quantify these relationships include Pearson correlation, Jaccard similarity, and cosine similarity.

Model-based approaches leverage predictive models to learn user-item interactions from historical data. Techniques such as matrix factorization, decision trees, Bayesian networks, and neural networks have been applied in this context \cite{Aggarwal2016Recommender}. A widely used matrix factorization technique is Singular Value Decomposition (SVD), which decomposes the user-item rating matrix into lower-dimensional representations known as latent factors. These latent factors capture hidden patterns in user preferences and item relationships, enhancing the prediction of unrated items.

In citation recommendation, a rating matrix is constructed from the citation graph, where citing and cited papers are treated as users and items, respectively. McNee et al. \cite{McNee2002RecommendingCitations} published the first study in this area, applying collaborative filtering techniques. They compared four algorithms and showed that user-item and item-item approaches outperformed traditional keyword-based methods. Sugiyama et al. \cite{Sugiyama2013Collab} used a paper-citation matrix and Pearson correlation to recommend citations based on similarity between citation vectors. Caragea et al. \cite{Caragea2013Collab} applied SVD to the citation graph, projecting papers into a latent space and outperforming traditional item-based approaches.

Although collaborative filtering methods can be effective, they often suffer from data sparsity and cold-start problems, which hinder performance when user-item interactions are limited. However, several studies have work towards addressing these issues. For example, Caragea et al. \cite{Caragea2013Collab} mitigated sparsity through matrix factorization, while Bansal et al. \cite{Bansal2016GRU} addressed the cold-start problem by leveraging gated recurrent units (GRUs) to analyze the textual content of papers. Moreover, many collaborative filtering methods, such as those by McNee et al. \cite{McNee2002RecommendingCitations}, Jia et al. \cite{Jia2017Analysis}, and Liu et al. \cite{Liu2015Context}, rely on the availability of seed citations in query papers. However, these may be unavailable in the early stages of research. Additionally, methods that primarily model relationships between entities may not adequately capture content-based similarities between items. As noted by Ali et al. \cite{Ali2021Review}, relatively few citation recommendation studies have utilized collaborative filtering techniques.

\subsubsection{Content-Based Citation Recommendation}
Content-based citation recommendation approaches analyze the textual features of papers to compute similarity and generate recommendations. Unlike collaborative filtering techniques, these methods are not affected by data sparsity issues and do not require query papers to contain partial citations \cite{Khusro2016Recommender}. Text representation learning involves transforming textual content into vector representations while preserving semantic and syntactic relationships. It is fundamental for natural language processing (NLP) tasks, including information retrieval and text classification. Text representation approaches can be broadly categorized into traditional bag-of-words (BoW) models, topic models, and deep learning methods \cite{Liu2023Text}.

BoW models represent documents as unordered collections of terms, treating each term as an independent feature. These models disregard word order and context, resulting in sparse, vocabulary-sized vector representations. He at al. \cite{He2010ContextAware} introduced the task of local citation recommendation and used a classic BoW model, TF-IDF (Term Frequency-Inverse Document Frequency) \cite{Salton1988TFIDF}, to represent citation contexts and compute similarity with candidate articles. Similarly, Tang et al. \cite{Tang2014Language} used TF-IDF to construct cross-language representations, while Makarov et al. \cite{Makarov2021Fusion} and Bhagavatula et al. \cite{Bhagavatula2018ContentBased} utilized TF-IDF and BM25 \cite{Robertson2009BM25}, respectively, as baseline methods in their citation recommendation works.

Topic modelling represents documents as probability distributions over latent topics, where each topic is itself a distribution over words. Unlike BoW models, topic models capture hidden semantic relationships by mapping documents to a lower-dimensional latent space. Popular methods such as Latent Dirichlet Allocation (LDA) \cite{Blei2003LDA} and Latent Semantic Analysis (LSA) \cite{Landauer1997LSA} have been applied in various citation recommendation studies. For example, Amami et al. \cite{Amami2016LDA} and Achakulvisut et al. \cite{Achakulvisut2016Fast} used LDA and LSA, respectively, to represent the abstracts of papers. However, as noted by F{\"a}rber et al. \cite{Farber2020Review}, topic modelling approaches have become less common in recent years.

More recently, deep learning-based text representation approaches have gained popularity. These methods use deep neural architectures to generate dense, low-dimensional embeddings that capture rich semantic relationships and mitigate sparsity \cite{Liu2023Text}. Bhagavatula et al. \cite{Bhagavatula2018ContentBased} proposed a two-phase model: NNSelect, which embeds papers using a supervised neural model, and NNRank, which re-ranks candidates with a feed-forward network. Their approach demonstrated significant improvements over the previous state-of-the-art ClusCite algorithm by Ren et al. \cite{Ren2014ClusCite}. Several studies have fine-tuned embedding models to learn citation-aware representations using citation-based training objectives. For example, Bhagavatula et al. \cite{Bhagavatula2018ContentBased} optimized a triplet loss function \cite{Wang2014Triplet} to align citing and cited papers closer together in the embedding space. Later, Cohan et al. \cite{Cohan2020SPECTER} introduced SPECTER, a transformer-based document embedding model initialized with SciBERT \cite{Beltagy2019SciBERT} and trained using a similar triplet loss approach. SPECTER outperformed previous distributed text representation models, including SciBERT and Doc2vec \cite{Le2014Doc2vec}, by encoding both semantic content and implicit citation relationships.

Although content-based methods are effective at generating recommendations based on textual similarity, they often overlook structural relationships between entities such as papers, authors, and venues. This lack of relational context reduces their ability to capture the broader factors that influence citation behavior, often leading to suboptimal recommendations. Addressing this gap is crucial, as citation practices are shaped not only by textual relevance but also by complex network dynamics. This limitation has consequently driven growing interest in hybrid approaches that integrate both textual and structural signals to produce more comprehensive and accurate citation recommendations.

\subsubsection{Graph-Based Citation Recommendation}
Graph-based citation recommendation approaches model relationships between entities within a network structure, enabling the capture of more nuanced interactions than traditional collaborative filtering and content-based approaches. These methods extract features from homogeneous or heterogeneous graphs. 

\begin{figure}[h]
    \centering
    \includegraphics[width=0.48\textwidth]{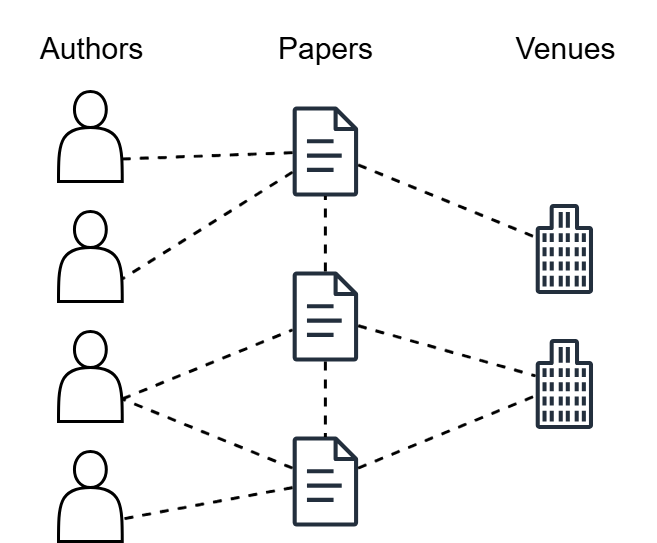}
    \caption{A heterogeneous bibliographic network consisting of author, paper, and venue nodes. Edges represent different semantic relations: authors write papers, papers cite other papers, and papers are published in venues.}
    \label{fig:heterogeneous_graph}
\end{figure}

Homogeneous networks contain only one type of node and one type of edge. For example, a citation network consists of nodes representing papers and edges representing citations. In contrast, heterogeneous networks consist of multiple types of entities (e.g., papers, authors, and venues) and relations (e.g., paper-paper citations, author-paper authorship), offering greater diversity and information, which often results in improved recommendations. An example of a heterogeneous bibliographic network is shown in Figure~\ref{fig:heterogeneous_graph}.

Traditional graph-based citation recommendation methods often used random walk techniques. A random walk is a Markov chain defined on a network, where each node corresponds to a state, and the walker moves between states according to transition probabilities. For example, Gori et al. \cite{Gori2006PaperRank} introduced PaperRank, an extension of PageRank \cite{Page1999PageRank}, which performs a biased random walk over citation networks to rank and recommend influential papers. Meng et al. \cite{Meng2013Unified} developed a personalized citation recommendation system based on a heterogeneous graph that integrated papers, authors, topics (extracted with LDA), and words, and applied an adapted random walk to produce recommendations. However, random walk-based methods can suffer from performance degradation when older, highly cited papers are overemphasized \cite{Son2017Multilevel}. To address this limitation, network representation learning techniques have become increasingly popular. 

Network representation learning approaches embed nodes in a graph into a low-dimensional vector space while preserving structural and semantic relationships. These embeddings enable similarity computations between candidate and query papers. Initial work in this area focused on homogeneous graphs. For instance, Gupta et al. \cite{Gupta2017TextAndGraph} and Cai et al. \cite{Cai2019Representation} applied DeepWalk \cite{Perozzi2014DeepWalk} and Node2vec \cite{Grover2016Node2vec}, respectively, to learn node embeddings for citation recommendation by performing random walks over homogeneous citation networks. More recently, these approaches have been extended to heterogeneous graphs. Ali et al. \cite{Ali2021MultiView} introduced a multi-view heterogeneous network embedding method that incorporates semantic relationships among papers, venues, authors, topics, and keywords to learn richer representations. Their method outperformed previous state-of-the-art approaches, including NNRank by Bhagavatula et al. \cite{Bhagavatula2018ContentBased}.

While graph-based methods are effective at modelling entity-level relationships and network dynamics, they often overlook the rich semantic context found in the textual content of papers. This limits their ability to capture topical relevance, which can lead to less contextually relevant recommendations. Hence, integrating structural and textual information is crucial to reflect the multifaceted nature of citation behavior.

\subsubsection{Hybrid Model-Based Citation Recommendation}
Given the complementary strengths and inherent limitations of collaborative filtering, content-based, and graph-based methods, recent work has increasingly leveraged hybrid approaches that integrate multiple information filtering strategies for improved citation recommendation.

Torres et al. \cite{Torres2004TechLens} introduced TechLens+, a hybrid citation recommendation system that combines a TF-IDF content-based method with a k-nearest neighbor collaborative filtering algorithm. By integrating and ranking citations from both methods, their system outperformed each technique when evaluated individually. Other works have incorporated graph-based global relevance as an additional ranking factor. For example, Bethard et al. \cite{Bethard2010PageRank} first generated a list of recommendations using a content-based approach, then re-ranked them based on factors such as PageRank scores, citation count, venue citations, author citations, and recency.

Recent studies have explored multi-view representation learning to enhance citation recommendation by combining textual content with network structure. Qiu et al. \cite{Qiu2021Integration} investigated probabilistic and concatenation fusion techniques to merge text and network embeddings, using the resulting unified representations to train a classifier for link prediction. Gupta et al. \cite{Gupta2017TextAndGraph} and Chanana et al. \cite{Chanana2019TextAndGraph} applied Canonical Correlation Analysis (CCA) \cite{Hardoon2004CCA} to learn linear transformations that project text and node embeddings into a shared subspace, which were then linearly combined. These studies highlighted that fused embeddings, which integrate multiple complementary information sources, outperform single-modality representations for citation recommendation.

Despite the strengths of multi-view representation learning, several challenges remain. Aligning heterogeneous modalities, such as textual and structural views, is difficult due to their differing semantic characteristics and statistical properties. Linear techniques such as CCA, while effective, are constrained in their ability to capture complex, non-linear relationships between data views \cite{Andrew2013DCCA}. These limitations highlight the need for more expressive models capable of learning richer cross-modal interactions for improved citation recommendation.

\subsection{Evaluation Methods}
Evaluation methods to assess the performance of citation recommendation systems are generally categorized into offline, online, and user study-based approaches \cite{Farber2020Review}. Offline evaluation is the most commonly used method, where system performance is measured against a predefined ground truth without user involvement. In contrast, online evaluation examines acceptability rates in deployed recommendation systems, and user studies assess user satisfaction through explicit ratings.

\subsubsection{Offline Evaluation}
\label{sec:offline-evaluation}
Offline evaluation of citation recommendation systems assesses the accuracy of recommendations by comparing them to the reference lists of query papers, which act as the ground truth. A recommendation is considered relevant if it matches one of the cited papers. The primary advantage of offline evaluation is its efficiency, as results can often be obtained in minutes or hours, unlike online evaluation and user studies, which may take weeks or months. However, offline evaluation is often criticized for emphasizing accuracy while ignoring user-centric factors \cite{Beel2015Review}. As a result, high performance in offline evaluations does not necessarily translate to user satisfaction in real-world scenarios.

Offline evaluation typically involves using standard information retrieval metrics to assess both the relevance and ranking quality of citation recommendations. The most commonly used metrics include precision, recall, Mean Average Precision (MAP), Mean Reciprocal Rank (MRR), and Normalized Discounted Cumulative Gain (nDCG) \cite{Ali2021Review}. Several studies, including those by Ali et al. \cite{Ali2021Review} and Lee et al. \cite{Lee2023TwentyYears}, recommend using a combination of accuracy-based metrics, such as precision and recall, and ranking-based metrics, such as MAP, MRR, and nDCG, to provide a more comprehensive evaluation.

Another notable criticism is the presence of citing biases. Using reference lists as ground truth can be problematic because the original citing behavior may be biased \cite{Farber2020Review}. Consequently, even if a citation recommendation system recommends equally valid or better papers, they will not be recognized as "correct" if they were not cited by the query paper. Offline evaluation thus ignores the cite-worthiness of recommendations and the relevance of alternative citations. Common forms of citing biases are listed in Table \ref{tab:citing-biases}.

\begin{table}[h]
  \centering
  \caption{Common forms of citing biases}
  \label{tab:citing-biases}
  \setlength{\tabcolsep}{12pt}
  \begin{tabular}{ll}
    \hline
    \textbf{Bias Type} & \textbf{Description} \\
    \hline
    Content bias & Favoring papers that align with the author's viewpoint. \\
    Author bias & Preferring works by well-known or familiar authors. \\
    Venue bias & Prioritizing publications from prestigious conferences or journals. \\
    Timeliness bias & Favoring newer or older, well-cited papers while ignoring less visible but relevant works. \\
    Access bias & Overlooking paywalled or less-publicized papers due to limited accessibility. \\
    \hline
  \end{tabular}
\end{table}

These biases represent limitations that should be considered when assessing the accuracy of citation recommendation systems. Despite these concerns, offline evaluation remains the predominant evaluation approach, with relatively few citation recommendation studies incorporating user feedback. Given its practicality and common use, our work focuses on offline evaluation.

\subsubsection{Online Evaluation}
Online evaluation measures the acceptance of recommendations in live systems, typically through click-through rates, which act as implicit indicators of user satisfaction \cite{Beel2015Review}. Unlike offline evaluations, online methods offer more practical insights into user behavior but are less scalable, as they require access to deployed systems. At the time of their review, Farber et al. \cite{Farber2020Review} noted that no citation recommendation studies had applied online evaluation methods. This still appears to be the case in more recent work.

\subsubsection{User Studies}
User studies measure recommendation quality through explicit user ratings and are often regarded as the most comprehensive evaluation method \cite{Beel2015Review}. However, they are resource-intensive and require a substantial number of participants to produce meaningful results. Consequently, only a limited number of citation recommendation works have incorporated human-subject evaluations. For example, Saier et al. \cite{Saier2020SemanticModelling} and McNee et al. \cite{McNee2002RecommendingCitations} complemented their offline evaluations with user studies to better assess the effectiveness of their models, while Khadka et al. \cite{Khadka2018CitationContext} relied solely on user studies for their evaluation.

\section{Proposed Approach}
\label{sec:proposed-approach}
Previous work on citation recommendation has demonstrated the effectiveness of multi-view representation learning for integrating textual and structural signals. However, existing methods that apply linear fusion techniques, such as CCA, are limited in their ability to capture complex, non-linear relationships between modalities. To address this, we propose a DCCA-based approach that projects text and node embeddings of scientific articles into a shared latent space using neural networks. Despite its success in other fields, DCCA has yet to be applied to citation recommendation tasks, which presents a promising direction for improvement. The following subsections detail our approach, including problem definition, representation learning architecture, embedding pipelines, and recommendation mechanism.

\subsection{Problem Definition}
Let $P = \{p_1,\dots,p_n\}$ be a set of training papers, each associated with a title, abstract, and a set of citations. The citation graph $G = (V, E)$ consists of nodes $V$, corresponding to the papers in $P$, and directed edges $E \subseteq V \times V$ representing citation links, where $(p_i, p_j) \in E$ if paper $p_i$ cites paper $p_j$. Each paper $p_i$ is described by two data views: its textual content $x_i$ and its structural context $y_i$. The objective is to learn an embedding function $f : p_i \mapsto \mathbb{R}^d$ that maps each paper into a shared latent space where both semantic and structural similarities are preserved. This space is then used to retrieve relevant citations for a given query paper.

\subsection{System Architecture}
\begin{figure}[h]
    \centering
    \includegraphics[width=1\textwidth]{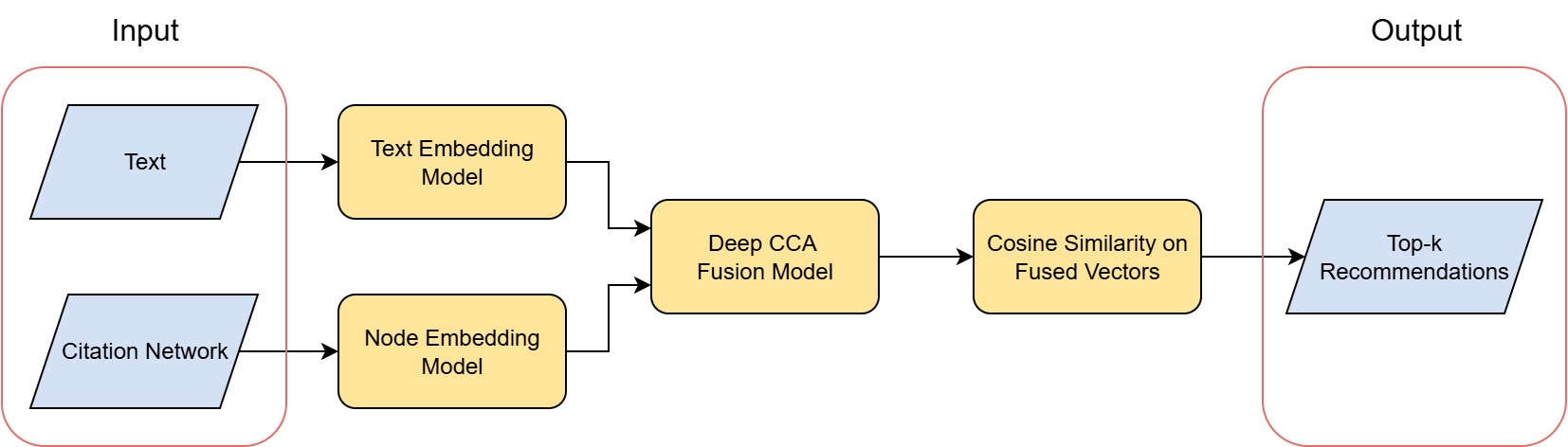}
    \caption{Architecture of the proposed citation recommendation system. Textual and structural features are encoded separately, then projected into a shared latent space using DCCA. Citation relevance is computed by measuring similarity in this space.}
    \label{fig:architecture}
\end{figure}

Figure \ref{fig:architecture} presents the architecture of our proposed citation recommendation system, which integrates both textual content and citation network information from scientific articles. Separate embedding models independently capture the unique features of each modality. These modality-specific embeddings are then projected into a shared latent space using DCCA, which models complex, non-linear relationships between the data views. Citation relevance is computed based on the similarity of representations in this joint space. This design preserves the distinct characteristics of each modality prior to fusion and leverages DCCA to address the limitations of linear multi-view approaches such as CCA, enabling more expressive representation learning.

\subsection{Embedding Generation Pipelines}
\begin{figure}[h]
    \centering
    \includegraphics[width=0.95\textwidth]{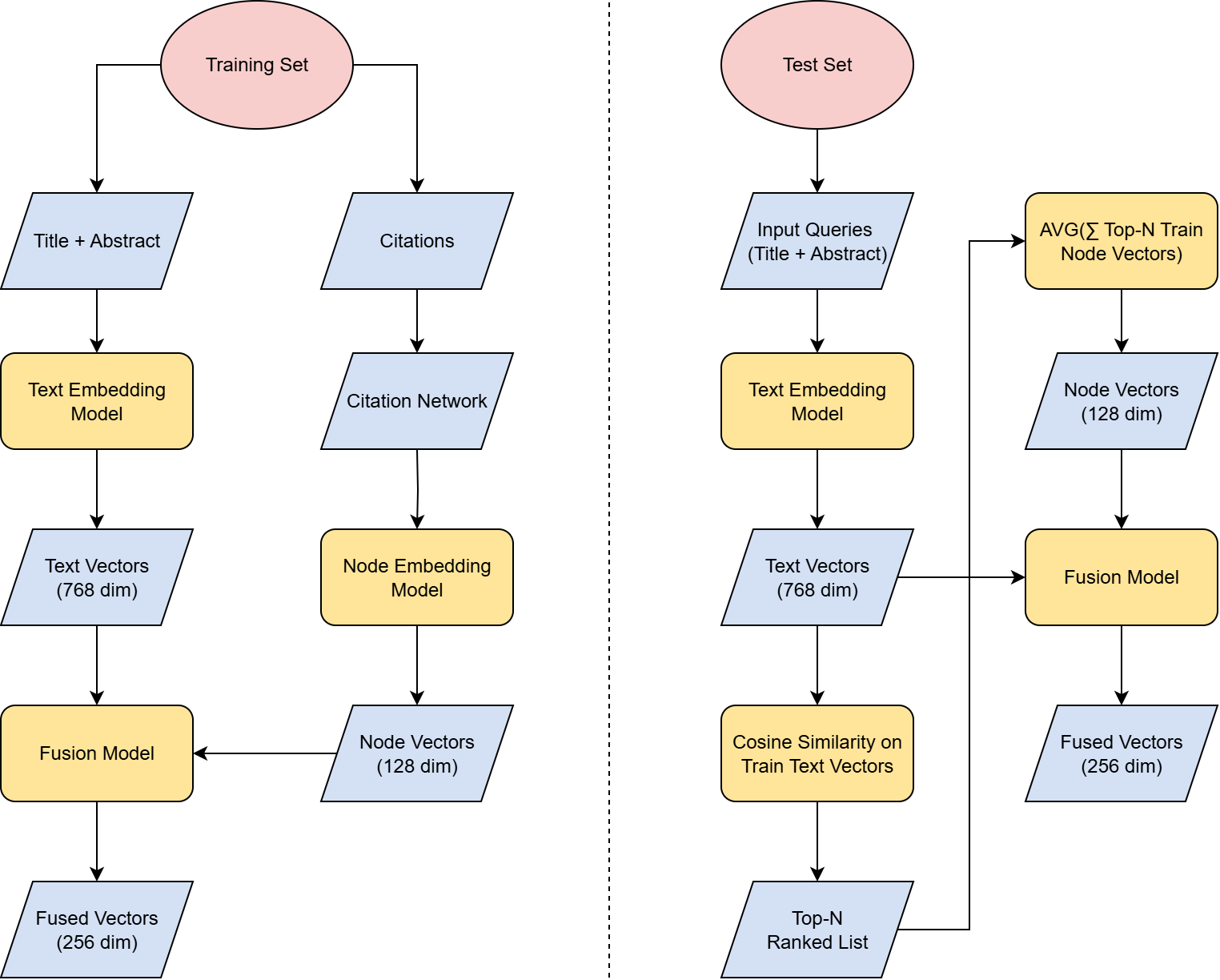}
    \caption{Overview of the pipelines for generating text, node, and fused embeddings during training and testing.}
    \label{fig:train_test_pipelines}
\end{figure}

Figure \ref{fig:train_test_pipelines} illustrates the pipelines used to generated text, node, and fused embeddings for both training and test data.

\subsubsection{Training Pipeline}
The training pipeline consists of the following steps:

{
\renewcommand{\labelitemi}{--}
\begin{itemize}
    \item \textbf{Text embeddings:} The title and abstract of each training paper are concatenated and passed through a text embedding model to generate a vector representation that captures the semantic content of the paper.
    
    \item \textbf{Node embeddings:} A citation network is constructed with training papers as nodes and citations as edges. A node embedding model is then applied to generate vector representations for each paper, capturing their structural relationships.

    \item \textbf{Fused embeddings:} The text and node embeddings are used to train a fusion model that projects both views into a shared latent space, enabling the system to leverage complementary information from both modalities and provide more comprehensive citation recommendations.
\end{itemize}
}

\subsubsection{Testing Pipeline}
The testing pipeline involves the following steps:

{
\renewcommand{\labelitemi}{--}
\begin{itemize}
    \item \textbf{Text embeddings:} As in training, the content of each test paper is passed through a text embedding model.

    \item \textbf{Node embeddings:} As the citation network is constructed using only training papers, test papers lack direct node embeddings. To estimate them, we average the node embeddings of the top-$N$ most textually similar training papers based on cosine similarity. This approximation ensures that the fusion model receives paired input views at inference time.
    
    \item \textbf{Fused embeddings:} The text and inferred node embeddings are passed through the trained fusion model to generate fused embeddings for each test paper.
\end{itemize}
}

\subsection{Text Representation}
Intuitively, the textual content of scientific articles provides a strong signal for identifying relevant citations, as papers within the same research field often share similar terminologies and discuss related concepts. We consider the content of each paper to be the concatenation of its title and abstract. We evaluate both traditional bag-of-words (BoW) models using TF-IDF and transformer-based text representation approaches, specifically SciBERT and SPECTER2.

\subsubsection{Term Frequency-Inverse Document Frequency (TF-IDF)}
TF-IDF, a classical BoW model \cite{Salton1988TFIDF}, is used as a baseline in our work. It assigns weights to terms based on two components: term frequency (TF), which reflects how often a term appears in a document, and inverse document frequency (IDF), which measures how rare the term is across the corpus. The product of TF and IDF produces a score that highlights terms that are frequent within a document but distinctive across the corpus. The TF-IDF score for a term $t$ in a document $d$ is defined as:

\begin{equation}
TF\text{-}IDF(t, d) = TF(t, d) \times \log\left(\frac{N}{df(t)}\right)
\end{equation}

where $TF(t,d)$ is the term frequency of term $t$ in document $d$, $N$ is the total number of documents in the collection, and $df(t)$ is the number of documents containing $t$. The logarithmic component, $\log\left(\frac{N}{df(t)}\right )$, represents the IDF.

Although TF-IDF is simple and effective for many tasks, it has limitations. As a BoW model, it ignores the order of terms in documents and the context in which they appear. This lack of contextual awareness may limit its ability to capture deep semantic relationships between terms.

\subsubsection{SciBERT}
SciBERT is a transformer-based language model designed for processing scientific text \cite{Beltagy2019SciBERT}. It builds upon the BERT (Bidirectional Encoder Representations from Transformers) architecture \cite{Devlin2019BERT}, but differs in its training corpus. While BERT is trained on general-domain corpora such as Wikipedia, SciBERT is trained on a large collection of full-text scientific papers from Semantic Scholar\footnote{\url{https://www.semanticscholar.org}}. This domain-specific training enhances its performance on scientific natural language processing (NLP) tasks. Unlike traditional BoW models, SciBERT generates dense, context-aware word embeddings using a transformer architecture. The encoding process begins with tokenization, which splits input text into individual tokens. Each token is mapped to an initial embedding from SciBERT's pre-trained vocabulary. These embeddings are then passed through a multi-layer transformer, where each layer consists of two main components: a self-attention mechanism and a feedforward neural network. 

The self-attention mechanism enables the model to compute relationships between tokens regardless of their position in the sequence. Unlike unidirectional models such as GPT (Generative Pre-trained Transformer) \cite{Radford2018GPT}, SciBERT operates bidirectionally, meaning each token's representation is influenced by both preceding and succeeding  tokens. Attention scores are calculated to determine how much influence each token should have on the representation of others. The embedding of each token is updated as a weighted sum of all other token embeddings, with the weights determined by the attention scores.

Following the self-attention step, the updated embeddings pass through a feedforward neural network that applies non-linear transformations to capture more complex relationships in the text. The final output of SciBERT is a set of dense, contextualized embeddings for each token in the input text. To obtain a single representation for an entire sequence, a common pooling strategy is to compute the mean of these token embeddings.

\subsubsection{SPECTER2}
SPECTER2 \cite{Singh2023SPECTER2} is a collection of citation-aware document embedding models that build upon the original SPECTER model \cite{Cohan2020SPECTER}. While general-purpose text representation methods focus solely on content, citation-informed text embeddings also implicitly integrate citation information to capture bibliographic relationships. This approach provides a more focused representation of the content of a paper. Like its predecessor, SPECTER2 is initialized with SciBERT and trained using a citation-based objective to refine the transformer's parameters. It optimizes a custom triplet loss function to align citing and cited papers closer together in the representation space:

\begin{equation}
\mathcal{L}_{\text{triplet}} = \max \left( d(P^{Q}, P^{+}) - d(P^{Q}, P^{-}) + m, 0 \right)
\end{equation}

where $P^Q$ is the embedding of a query paper $Q$, $P^+$ is the embedding of a positive paper (cited by $Q$), $P^-$ is the embedding of a negative paper (not cited by $Q$), the function $d$ denotes the Euclidean distance function, and $m$ is a margin hyperparameter.

SPECTER2 improves upon SPECTER by significantly increasing the size and diversity of its training data and by introducing adapter models for retrieval, classification, regression, and ad-hoc search tasks. SPECTER was trained solely on citation prediction, with approximately 70\% of its training data derived from computer science and biomedical fields. This narrow focus limits its effectiveness in other domains and tasks. In contrast, SPECTER2 is trained on six million triplets, 10 times more than SPECTER, and spans 23 different fields of study. Its adapter models further enhance generalization across various downstream applications. Adapters are linear modules attached to each transformer layer \cite{Houlsby2019Transfer}, and during fine-tuning, only their parameters are updated, while the base transformer remains fixed. Each task uses a suitable loss function: cross-entropy for classification, mean squared error for regression, and triplet margin loss for retrieval and search. We use the retrieval adapter, as it is fine-tuned for proximity-based search tasks.

SPECTER2 was trained using up to 10 triplet instances per query paper in the training set. Positive candidates were selected from the direct citations of each query paper. Six easy negatives were sampled based on the field of study (four from the same field as the query and two from different fields), while four hard negatives were chosen from papers cited by the query paper's citations. The presence of hard negatives improves performance by providing a more nuanced training signal, as they are semantically related to the query paper despite not being directly cited.

\subsection{Network Representation}
\begin{figure}[h]
    \centering
    \includegraphics[width=0.32\textwidth]{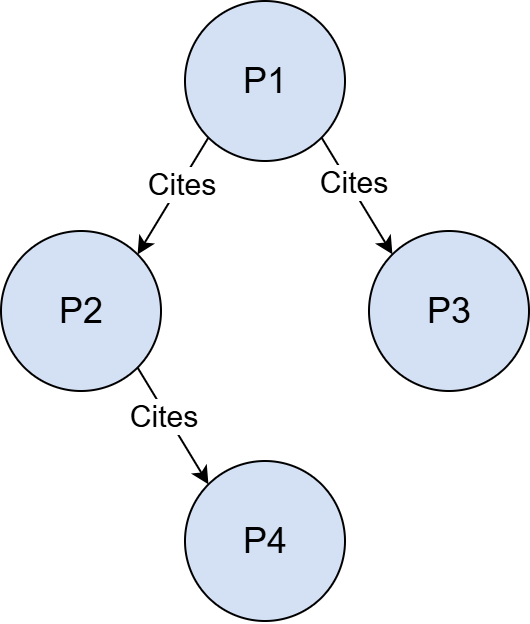}
    \caption{A directed citation network where each node corresponds to a paper and each edge indicates a citation.}
    \label{fig:graph}
\end{figure}

Citation networks capture meaningful relationships between papers, as those within the same research field tend to cite each other frequently. We evaluate two node embedding methods, DeepWalk and Node2vec, on a directed, homogeneous citation network, where nodes represent papers and edges denote citation links, as depicted in Figure \ref{fig:graph}.

\subsubsection{DeepWalk}
DeepWalk learns node embeddings by performing uniform random walks on a graph \cite{Perozzi2014DeepWalk}. Starting from a given node, it repeatedly transitions to a randomly chosen neighbor for a fixed walk length, generating sequences of nodes. These sequences are treated as \textit{sentences} in natural language, where each node is considered a \textit{word} and its neighboring nodes within the walk serve as its \textit{context}. This process effectively captures the local structural relationships between nodes.

The Skip-Gram model from Word2vec \cite{Mikolov2013Word2vec} is then trained on these sequences to learn node embeddings by predicting the context of each node within a fixed window size. Skip-Gram is a shallow neural network consisting of an input layer and an output layer, connected by an embedding (hidden) layer. The input is a one-hot encoded vector representing a node, and the output layer applies a softmax function to estimate a probability distribution over all nodes, indicating the likelihood of each node being within the context of the input node. The model is trained to minimize the cross-entropy loss between the predicted distribution and the actual context nodes. After training, the embedding of each node is obtained from the learned weights between the input and hidden layers.

\subsubsection{Node2vec}
Node2vec extends DeepWalk by using biased random walks for more flexible graph exploration \cite{Grover2016Node2vec}. Instead of uniform random walks, Node2vec performs second-order biased walks, where the transition probability depends on both the current node and the previous node. Two hyperparameters, $p$ (return parameter) and $q$ (in-out parameter), control the bias of the walk, allowing for a balance between local and global exploration of the graph.

The parameter $p$ influences the likelihood of revisiting the previous node: a low $p$ increases the chance of backtracking, encouraging local, inward-biased walks, while a high $p$ discourages backtracking, promoting outward exploration. The parameter $q$ affects whether the walk stays close to the previous neighborhood or moves further away: a low $q$ favors moving away, approximating depth-first search behavior, whereas a high $q$ encourages remaining within the neighborhood, resembling breadth-first search. When both $p=1$ and $q=1$, the walk reduces to an unbiased uniform random walk, equivalent to that used in DeepWalk.

\subsection{Feature Fusion}
Relying on a single information filtering approach has its limitations: text representation learning ignores structural relationships between papers, while network-based representations overlook rich content features. Although we learn text and node embeddings separately, combining them into a unified representation space through multi-view representation learning can address these shortcomings and improve recommendation quality by leveraging complementary information from both modalities. We propose using DCCA to integrate the text and node embeddings of scientific articles and compare its performance with the baseline CCA method.

\subsubsection{Canonical Correlation Analysis (CCA)}
CCA is a statistical method that learns linear transformations for two sets of variables such that their projected representations are maximally correlated in a shared subspace \cite{Hardoon2004CCA}. Formally, let $\mathbf{X} = [\mathbf{x}_1, \mathbf{x}_2, \ldots, \mathbf{x}_n] \in \mathbb{R}^{d_x \times n}$ and $\mathbf{Y} = [\mathbf{y}_1, \mathbf{y}_2, \ldots, \mathbf{y}_n] \in \mathbb{R}^{d_y \times n}$ denote two views (e.g., text and node embeddings) with $n$ samples and feature dimensions $d_x$ and $d_y$, respectively. CCA finds projection matrices $\mathbf{W}_x \in \mathbb{R}^{d_x \times d}$ and $\mathbf{W}_y \in \mathbb{R}^{d_y \times d}$ that maximize the correlation between the projected views in a $d$-dimensional space:
\begin{align}
(\mathbf{W}_x, \mathbf{W}_y) &= \operatorname*{argmax}_{\mathbf{W}_x, \mathbf{W}_y} \; \text{corr}(\mathbf{W}_x^\top \mathbf{X}, \mathbf{W}_y^\top \mathbf{Y}) \\
&= \operatorname*{argmax}_{\mathbf{W}_x, \mathbf{W}_y} \sum_{j=1}^d 
\frac{
\operatorname{Cov}\left(\mathbf{w}_x^{(j)\top} \mathbf{X},\ \mathbf{w}_y^{(j)\top} \mathbf{Y}\right)
}{
\sqrt{
\operatorname{Var}\left(\mathbf{w}_x^{(j)\top} \mathbf{X}\right) \cdot \operatorname{Var}\left(\mathbf{w}_y^{(j)\top} \mathbf{Y}\right)
}
} \\
&= \operatorname*{argmax}_{\mathbf{W}_x, \mathbf{W}_y} \sum_{j=1}^d \rho_j
\end{align}

where $\mathbf{w}_x^{(j)}$ and $\mathbf{w}_y^{(j)}$ denote the $j$-th columns of $\mathbf{W_x}$ and $\mathbf{W_y}$, respectively, the vectors $\mathbf{x}_j' = \mathbf{w}_x^{(j)\top} \mathbf{X}$ and $\mathbf{y}_j' = \mathbf{w}_y^{(j)\top} \mathbf{Y}$ represent the $j$-th pair of canonical variables, and $\rho_j$ denotes the $j$-th canonical correlation.

Although CCA has been shown to be effective for citation recommendation \cite{Gupta2017TextAndGraph, Chanana2019TextAndGraph}, its linearity limits its ability to capture complex, non-linear relationships between modalities \cite{Andrew2013DCCA}.

\subsubsection{Deep Canonical Correlation Analysis (DCCA)}
\begin{figure}[h]
    \centering
    \includegraphics[width=0.75\textwidth]{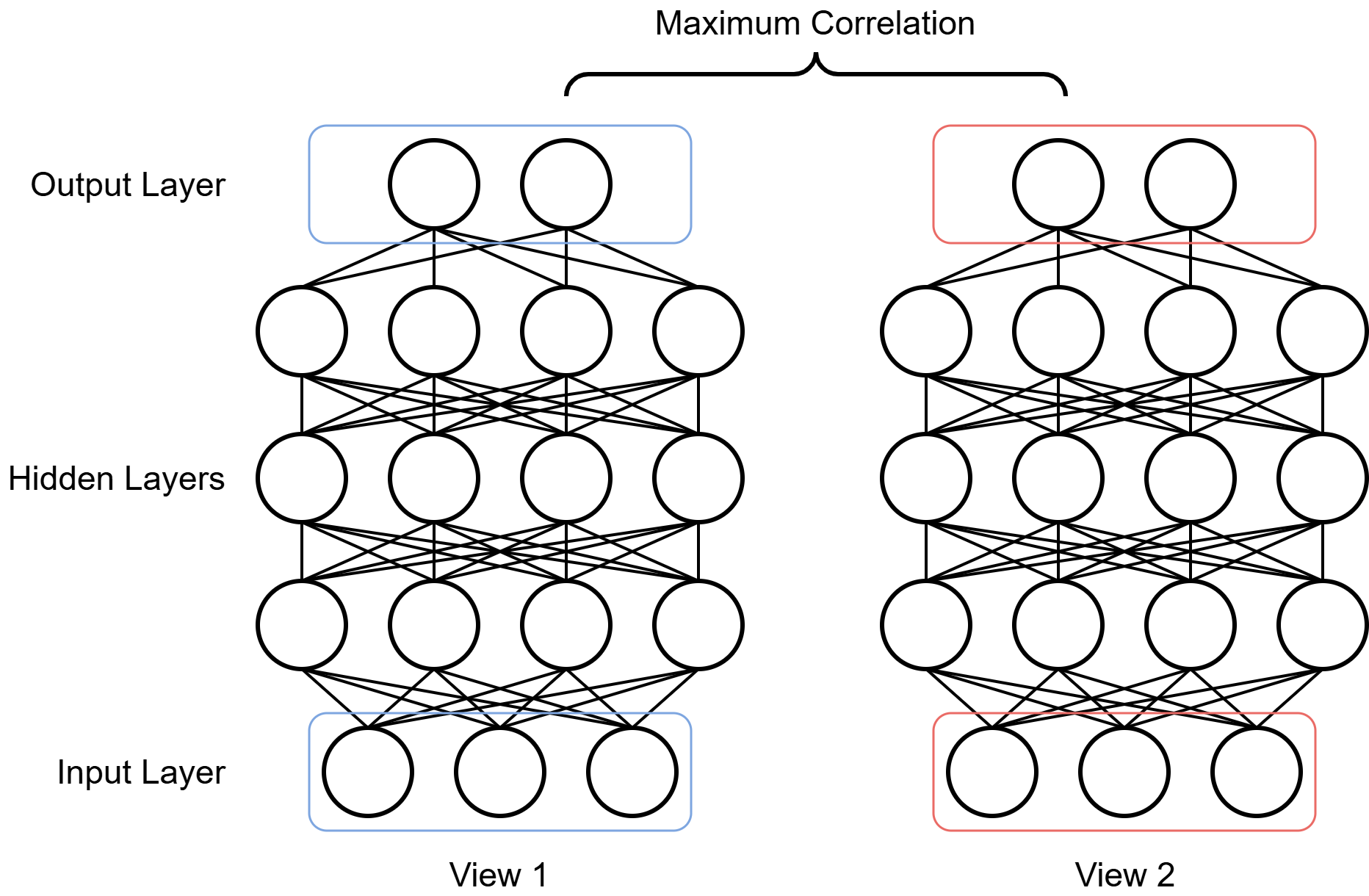}
    \caption{Generalized DCCA architecture, consisting of two fully connected deep neural networks.}
    \label{fig:dcca}
\end{figure}

DCCA extends CCA by replacing linear projections with non-linear transformations learned by neural networks, enabling the capture of more complex representations \cite{Andrew2013DCCA}. Each data view is passed through a separate multilayer perceptron (MLP), with activation functions (e.g., ReLU, sigmoid) applied at each neuron to introduce non-linearities. The two networks are trained jointly to maximize the correlation between their output representations. A generalized DCCA architecture is illustrated in Figure \ref{fig:dcca}. It consists of two fully connected deep neural networks, one for each data view, that map input vectors into a shared $d$-dimensional space where their representations are maximally correlated.

Formally, let $f_x(\mathbf{X}; \theta_x)$ and $f_y(\mathbf{Y}; \theta_y)$ denote neural networks parameterized by $\theta_x$ and $\theta_y$, respectively. These networks output transformed representations $\mathbf{X}' \in \mathbb{R}^{n \times d}$ and $\mathbf{Y}' \in \mathbb{R}^{n \times d}$. The shared loss function minimizes the negative sum of the canonical correlations $\rho_j$ between the transformed outputs of the two views:
\begin{equation}
\mathcal{L}_{\text{DCCA}} = - \sum_{j=1}^{d} \rho_j
\end{equation}

During training, the gradients of this objective are computed and used to jointly update the parameters of both neural networks via backpropagation.

\subsubsection{Fusion Strategies}
Although DCCA projects both the textual and structural views into a shared latent space where correlation is maximized, combining these aligned embeddings into single representation that effectively captures complementary information from both modalities still requires an appropriate fusion method. We evaluate three strategies for performing this fusion:

\begin{enumerate}
    \item \textbf{Simple concatenation:} As a baseline, we concatenate the original, unprojected text and node embeddings. This straightforward approach does not utilize any form of alignment or correlation maximization, which may limit its ability to integrate information across modalities.

    \item \textbf{Projected concatenation:} For both CCA and DCCA, we concatenate their projected embeddings, which preserves the correlation-aligned features of each modality separately and results in fused embeddings with dimensionality $2d$.
    
    \item \textbf{Linear combination:} Following the approach of Gupta et al. \cite{Gupta2017TextAndGraph} and Chanana et al. \cite{Chanana2019TextAndGraph}, we compute a weighted sum of the projected embeddings, controlled by a coefficient $\alpha \in [0, 1]$, defined as:

    \begin{equation}
    \mathbf{Z} = \alpha \cdot \mathbf{X}' + (1 - \alpha) \cdot \mathbf{Y}'
    \end{equation}
    
    where $\mathbf{X}'$ and $\mathbf{Y}'$ are the projected text and node embeddings, respectively. This strategy preserves the dimensionality of the projected representation space $d$.
\end{enumerate}

\subsection{Ranking Recommendations}
After fusing the text and node embeddings into a shared latent space, citation recommendation is performed by ranking candidate papers based on their similarity to a given query paper within this space. Specifically, we compute the cosine similarity between the fused embedding of the query paper and those of the papers in the training corpus. The top-$k$ most similar papers are then returned as citation recommendations. The cosine similarity between two vectors $A$ and $B$ is defined as:
\begin{equation}
\cos(\theta) = \frac{A \cdot B}{\|A\| \|B\|}
\end{equation}

where \( A \cdot B \) is the dot product of \( A \) and \( B \), and \( \|A\| \) and \( \|B\| \) are their respective $L_2$ norms.

The effectiveness of this retrieval phase directly depends on the quality of the fused embeddings: those that better capture complementary semantic and structural signals enable a more accurate identification of relevant citations.

\section{Experimental Setup}
\label{sec:experimental-setup}
In this section, we provide details of the experimental setup used to evaluate our citation recommendation model. We begin by describing the dataset, followed by the evaluation metrics and experimental settings.

\subsection{Dataset}
\begin{table*}[h]
  \caption{Summary of the DBLPv10 dataset used in our experiments. The table reports the publication span, number of papers, citation links, and average number of citations per paper across the original data, pruned subset, and the training and test sets}
  \label{tab:dataset-stats}
  \centering
  \begin{tabular}{lcccc}
    \toprule
    \textbf{Dataset} & \textbf{Years} & \textbf{Papers} & \textbf{Citations} & \textbf{Avg. Citations}\\
    \midrule
    Original & 1936-2018 & 3,079,007 & 25,166,994 & 8.17 \\
    Post-Pruning & 1968-2017 & 41,698 & 247,769 & 5.94 \\
    Training Set & 1968-2013 & 36,520 & 189,741 & 5.2 \\
    Test Set & 2014-2017 & 5,043 & 46,448 & 9.21 \\
    \bottomrule
  \end{tabular}
\end{table*}

We conducted our experiments using the DBLPv10\footnote{\url{https://www.aminer.cn/citation}} (Digital Bibliography \& Library Project) citation network dataset from ArnetMiner \cite{Tang2008ArnetMiner}, which includes metadata on computer science publications from 1936 to 2018. Each paper includes a unique identifier, title, abstract, publication year, authors, venue, and references. The full dataset comprises 3,079,007 papers and 25,166,994 citation relationships, with an average of 8.17 citations per paper. Following preprocessing steps similar to Gupta et al. \cite{Gupta2017TextAndGraph} and Bhagavatula et al. \cite{Bhagavatula2018ContentBased}, we filtered out papers with incomplete metadata, fewer than 15 incoming citations, or fewer than 20 outgoing references. After pruning, the dataset was reduced to 41,698 papers and 247,769 citation relationships, with an average of 5.94 citations per paper. Table \ref{tab:dataset-stats} summarizes the dataset statistics.

We used a temporal split for evaluation to ensure that recommendations for test papers only include citations to earlier works, which is more realistic than conventional random splits. The training set consists of papers published between 1968 and 2013, while the test set includes papers from 2014 to 2017. We retained only those test papers that cite at least one paper from the training set. The final training set comprises 36,520 papers and 189,741 citation relationships, with an average of 5.2 citations per paper. The test set consists of 5,043 papers and 46,448 citation relationships (approximately 24.5\% of all citations), with an average of 9.21 citations per paper. This higher average reflects the fact that test papers were required to cite at least one paper from the training set. 

\subsection{Evaluation Metrics}
For evaluation, the ground truth for each test paper is defined as its citations to papers in the training set, and a recommendation is considered relevant if it matches one of these citations. Top-$k$ thresholds are commonly used in recommendation tasks to focus evaluation on the most relevant recommendations. For global citation recommendation, Farber et al. \cite{Farber2020Review} recommends setting $k$ based on the average number of citations per paper rather than choosing arbitrarily large values (e.g., $k=50$), which may not reflect realistic citation behavior. Since our test set has an average of 9.21 citations per paper, we use $k=10$ as the smallest threshold and additionally evaluate at $k=15$ and $k=20$.

As discussed in Section \ref{sec:offline-evaluation}, several studies recommend using a combination of accuracy-based and ranking-based metrics for the offline evaluation of citation recommendation systems. This ensures both the relevance and ranking quality of recommendations are effectively evaluated. We assess recommendation performance using three standard information retrieval metrics: Precision@k, Recall@k, and Mean Average Precision@k (MAP@k). Precision and recall are selected to measure accuracy because of their widespread use in the research field. To assess ranking quality, we chose MAP instead of alternatives such as Mean Reciprocal Rank (MRR). While MRR evaluates the position of the first relevant recommendation, MAP considers all relevant items, providing a more comprehensive measure for global citation recommendation.

\subsubsection{Precision@k (P@k)}
P@k measures the proportion of relevant papers among the top-$k$ recommendations. It is defined as:
\begin{equation}
P@k = \frac{1}{|Q|} \sum_{i=1}^{|Q|} \frac{|R_i \cap T_i|}{k}
\end{equation}

where $Q$ is the set of test queries, $R_i$ is the set of ground truth citations for query $i$, and $T_i$ is the set of top-$k$ recommendations for query $i$.

\subsubsection{Recall@k (R@k)}
R@k measures the proportion of relevant papers retrieved in the top-$k$ recommendations. It is defined as:
\begin{equation}
R@k = \frac{1}{|Q|} \sum_{i=1}^{|Q|} \frac{|R_i \cap T_i|}{|R_i|}
\end{equation}

\subsubsection{Mean Average Precision@k (MAP@k)}
MAP@k is the mean of the Average Precision (AP) scores over all test queries. AP@k measures ranking quality by averaging the precision values at the positions where relevant papers appear within the top-$k$ recommendations. For a single query $i$, AP@k is defined as:
\begin{equation}
AP@k(Q_i) = \frac{1}{|R_i|} \sum_{j=1}^{k} P@j \cdot \text{rel}(j)
\end{equation}

where $P@j$ is the precision at position $j$, and $\text{rel}(j)$ is 1 if the recommendation at position $j$ is relevant, and 0 otherwise. MAP@k is then defined as:
\begin{equation}
MAP@k = \frac{1}{|Q|} \sum_{i=1}^{|Q|} AP@k(Q_i)
\end{equation}

\subsection{Experimental Settings}
\subsubsection{Text Representation}
For TF-IDF, we preprocess text to reduce noise and emphasize informative terms by removing non-alphabetic characters, converting to lowercase, removing stop words, and filtering out rare terms that occur fewer than five times in the training corpus. This results in a vocabulary that yields 16,340-dimensional representations. A logarithmic transformation is also applied to term frequencies to reduce sensitivity to document length.

The SciBERT and SPECTER2 embeddings are 768-dimensional, consistent with the underlying BERT architecture. SciBERT embeddings are generated by mean-pooling token embeddings from the transformer's final hidden layer, whereas SPECTER2 produces document-level embeddings directly without requiring token pooling post-processing. As a contextual language model, SciBERT captures semantic and syntactic relationships without requiring traditional text preprocessing steps such as stop-word or punctuation removal. For SPECTER2, we also use the retrieval adapter model, which is fine-tuned for proximity-based search tasks.

\subsubsection{Network Representation}
\label{sec:network-parameters}
The DeepWalk and Node2vec embeddings are 128-dimensional. We use 200 random walks per node, each with a walk length of 80 and a Skip-Gram window size of 10. For Node2vec, the return parameter is set to $p=4$ and the in-out parameter to $q=2$. These values were selected from a grid search over $p, q \in \{0.25, 0.5, 1, 2, 4\}$, following Grover et al. \cite{Grover2016Node2vec}, to explore different traversal biases.

The citation network is constructed using only training papers. For each test paper, we estimate its node embedding by averaging the embeddings of its top-$5$ most textually similar training papers, based on the cosine similarity of their textual representations.

\subsubsection{Feature Fusion}
For CCA, we learn a 128-dimensional shared subspace. For DCCA, we use shallow neural networks for both textual and structural views, as these proved to be the most effective during tuning. Each network is fully connected and consists of a single hidden layer with 128 neurons and a sigmoid activation function, followed by a 128-dimensional output layer. DCCA models are also trained for 20 epochs with a batch size of 256.

We evaluate three strategies for integrating text and node embeddings. The first strategy simply concatenates unprojected embeddings, resulting in 896-dimensional fused vectors. The second strategy concatenates projected CCA and DCCA embeddings, yielding 256-dimensional representations. Finally, the third strategy computes a linear combination of projected embeddings, preserving the 128-dimensional size, and we experiment with weighting coefficients $\alpha \in \{0.1, 0.25, 0.5, 0.75, 0.9\}$ to control the relative contributions of the textual and structural views.

\section{Results and Analysis}
\label{sec:results}
In this section, we present and discuss the experimental results of our citation recommendation system. We evaluate various text representation methods, network representation techniques, and feature fusion strategies. We hypothesize that DCCA will outperform the baseline CCA approach due to its ability to learn more expressive latent representations by capturing non-linear relationships between textual and structural views, which linear methods like CCA fail to capture. To assess the effectiveness of each approach, we report standard information retrieval metrics: Precision@k, Recall@k, and MAP@k, where $k \in \{10,15,20\}$. These metrics provide a comprehensive evaluation of both the relevance and ranking quality of the recommended citations.

\subsection{Text Representation}
This subsection evaluates the quality of citation recommendations generated using only textual content. We compare the traditional BoW model TF-IDF with two transformer-based embedding methods: SciBERT, a general-purpose scientific language model, and SPECTER2, a citation-aware document embedding model.

\begin{table*}[h]
    \caption{Text representation results}
    \label{tab:text-results}
    \centering
    \begin{tabular}{lccc ccc ccc}
        \toprule
        \textbf{Model} & \multicolumn{3}{c}{\textbf{Precision}} & \multicolumn{3}{c}{\textbf{Recall}} & \multicolumn{3}{c}{\textbf{Mean Average Precision}} \\
        \cmidrule(r){2-4} \cmidrule(r){5-7} \cmidrule(r){8-10}
        & P@10 & P@15 & P@20 & R@10 & R@15 & R@20 & MAP@10 & MAP@15 & MAP@20 \\
        \midrule
        TF-IDF & 0.1746 & 0.1446 & 0.1245 & 0.2256 & 0.2722 & 0.3067 & \textbf{0.138} & \textbf{0.1493} & \textbf{0.1562} \\
        SciBERT & 0.0872 & 0.0724 & 0.0632 & 0.1154 & 0.1396 & 0.1604 & 0.0621 & 0.0662 & 0.0689 \\
        SPECTER2 & \textbf{0.1775} & \textbf{0.1466} & \textbf{0.1268} & \textbf{0.2297} & \textbf{0.2771} & \textbf{0.3156} & 0.1354 & 0.1467 & 0.1538 \\
        \bottomrule
    \end{tabular}
\end{table*}

Table \ref{tab:text-results} presents the results of the text representation methods. Notably, SPECTER2, which is contextualized and citation-aware, only slightly outperforms TF-IDF in terms of precision and recall, while marginally underperforming in MAP. Furthermore, TF-IDF consistently outperforms SciBERT, suggesting that well-tuned BoW models can remain competitive despite their relative simplicity. This aligns with the findings of Gupta et al. \cite{Gupta2017TextAndGraph}, who demonstrated that traditional BoW models, namely TF-IDF and BM25, can outperform Doc2vec, a distributed text representation model, in the task of citation recommendation. One possible explanation for TF-IDF's strong performance in our experiments is that it is directly fitted to the dataset, potentially benefiting from domain-specific knowledge. In contrast, both SciBERT and SPECTER2 are pre-trained on external corpora, which may limit their ability to adapt fully to this dataset. However, SciBERT and SPECTER2 offer a clear advantage in vector dimensionality, each using only 768 dimensions compared to TF-IDF's 16,340. This leads to substantial savings in memory usage and computational cost. 

Between the two transformer-based models, SPECTER2 significantly outperforms SciBERT across all metrics, supporting the expectation that citation-aware fine-tuning yields richer representations than those produced from general-purpose scientific language models.

\subsection{Network Representation}
This subsection evaluates citation recommendation performance based solely on network structure information. We compare two node embedding methods, DeepWalk and Node2vec, each applied on a homogeneous citation graph with three different text representation techniques, TF-IDF, SciBERT, and SPECTER2, to estimate embeddings for test nodes.

\begin{table*}[h]
    \caption{Network representation results. TF, SB, and S2 denote TF-IDF, SciBERT and SPECTER2, respectively, as used to find nearest neighbors for estimating test node embeddings}
    \label{tab:network-results}
    \centering
    \begin{tabular}{lccc ccc ccc}
        \toprule
        \textbf{Model} & \multicolumn{3}{c}{\textbf{Precision}} & \multicolumn{3}{c}{\textbf{Recall}} & \multicolumn{3}{c}{\textbf{Mean Average Precision}} \\
        \cmidrule(r){2-4} \cmidrule(r){5-7} \cmidrule(r){8-10}
        & P@10 & P@15 & P@20 & R@10 & R@15 & R@20 & MAP@10 & MAP@15 & MAP@20 \\
        \midrule
        $\text{DeepWalk}_\text{TF}$ & 0.1399 & 0.1187 & 0.1043 & 0.165 & 0.2053 & 0.2362 & 0.0929 & 0.1029 & 0.1094 \\
        $\text{DeepWalk}_\text{SB}$ & 0.0779 & 0.0687 & 0.0623 & 0.0904 & 0.118 & 0.1411 & 0.045 & 0.0508 & 0.0547 \\
        \textbf{$\text{DeepWalk}_\text{S2}$} & \textbf{0.1444} & \textbf{0.1229} & \textbf{0.1081} & \textbf{0.1715} & \textbf{0.2134} & \textbf{0.2467} & \textbf{0.0939} & \textbf{0.1045} & \textbf{0.1112} \\
        $\text{Node2vec}_\text{TF}$ & 0.1392 & 0.1182 & 0.104 & 0.1645 & 0.205 & 0.2363 & 0.0929 & 0.1031 & 0.1095 \\
        $\text{Node2vec}_\text{SB}$ & 0.0783 & 0.0688 & 0.0621 & 0.0909 & 0.1189 & 0.1412 & 0.0449 & 0.0509 & 0.0547 \\
        $\text{Node2vec}_\text{S2}$ & 0.1429 & 0.1222 & 0.1075 & 0.1686 & 0.2122 & 0.2453 & 0.0933 & 0.1042 & 0.1109 \\
        \bottomrule
    \end{tabular}
\end{table*}

Table \ref{tab:network-results} summarizes the performance of the network representation methods. Among all configurations, similarity based on SPECTER2 consistently produces the most effective results, highlighting the strength of SPECTER2's citation-informed embeddings. In terms of network representation models, DeepWalk slightly outperforms Node2vec overall. This indicates that despite relying on a simpler uniform random walk mechanism compared to Node2vec's biased random walks, DeepWalk still effectively captures structural information. Although these network-based methods underperform compared to the text-based approaches, they still offer reasonable accuracy and serve as valuable complementary signals for citation recommendation.

As discussed in Section \ref{sec:network-parameters}, Node2vec's hyperparameters ($p=4$, $q=2$) were selected via a grid search over $p,q \in \{0.25,0.5,1,2,4\}$ to explore varying traversal biases. A high return parameter, $p$, discourages backtracking, promoting outward exploration, while a moderate in-out parameter, $q$, encourages walks to stay within the local neighborhood of the previous node. This combination suggests that an effective balance between global exploration and preservation of local structures is optimal for capturing patterns in citation networks.

\subsection{Feature Fusion}
The motivation for feature fusion is that while text and node embeddings are learned independently, combining them into a unified representation may improve recommendation quality by leveraging complementary information from both data views. We evaluate a baseline fusion strategy that simply concatenates the original, unprojected embeddings and compare the performance of DCCA with CCA.

\subsubsection{Simple Concatenation}
\begin{table*}[h]
    \caption{Feature fusion results using simple concatenation of unprojected text and node embeddings}
    \label{tab:simple-concat}
    \centering
    \resizebox{\textwidth}{!}{
    \begin{tabular}{llccccccccc}
        \toprule
        \parbox{1.2cm}{\textbf{Text Model}} & \parbox{1.2cm}{\textbf{Node Model}} & \multicolumn{3}{c}{\textbf{Precision}} & \multicolumn{3}{c}{\textbf{Recall}} & \multicolumn{3}{c}{\textbf{Mean Average Precision}} \\
        \cmidrule(r){3-5} \cmidrule(r){6-8} \cmidrule(r){9-11}
        & & P@10 & P@15 & P@20 & R@10 & R@15 & R@20 & MAP@10 & MAP@15 & MAP@20 \\
        \midrule
        SciBERT & DeepWalk & 0.0936 & 0.0805 & 0.0721 & 0.1125 & 0.1424 & 0.1681 & 0.0581 & 0.0645 & 0.0691 \\
        SciBERT & Node2vec & 0.0937 & 0.0804 & 0.0723 & 0.1128 & 0.1418 & 0.1676 & 0.0587 & 0.065 & 0.0696 \\
        SPECTER2 & DeepWalk & 0.1684 & \textbf{0.1421} & \textbf{0.1236} & \textbf{0.2073} & \textbf{0.2538} & \textbf{0.2887} & 0.1201 & 0.1325 & 0.14 \\
        SPECTER2 & Node2vec & \textbf{0.1685} & 0.1412 & 0.1229 & 0.2063 & 0.2524 & 0.2874 & \textbf{0.1213} & \textbf{0.1333} & \textbf{0.1408} \\
        \bottomrule
    \end{tabular}
    }
\end{table*}

Table \ref{tab:simple-concat} presents the results of the baseline fusion strategy that simply concatenates unprojected text and node embeddings. This method acts as a reference point for evaluating more advanced integration techniques. Compared to single-view representations, simple concatenation improves performance over standalone DeepWalk and Node2vec embeddings. Notably, Node2vec consistently achieves slightly stronger MAP scores than DeepWalk across all fusion configurations, indicating that it may be better suited for integration in fusion-based approaches. Compared to text-only models, the effectiveness of simple concatenation is mixed. For SciBERT, fusion consistently improves precision and generally enhances recall. However, MAP scores are typically higher with the text-only representation, suggesting that while simple concatenation helps identify relevant citations, it may negatively impact ranking quality. In contrast, for SPECTER2, the fused representations perform worse across all metrics.

Overall, while simple concatenation increases representational capacity by combining the two modalities, it may also introduce redundancy or noise, diminishing its effectiveness. These findings highlight the need for more sophisticated fusion strategies that better align and balance the contributions of each modality.

\subsubsection{Projected Concatenation and Linear Combination}
\begin{table*}[h]
    \caption{Feature fusion results using concatenation of projected CCA and DCCA embeddings. SB and S2 denote SciBERT and SPECTER2, respectively, while DW and N2 refer to DeepWalk and Node2vec. Relative gains (\%) are computed by comparing the top-performing DCCA model ($\text{DCCA}_\text{S2+N2}$) against the corresponding CCA model ($\text{CCA}_\text{S2+N2}$) for each metric}
    \label{tab:projection-concat-results}
    \centering
    \resizebox{\textwidth}{!}{
    \begin{tabular}{lccc ccc ccc}
        \toprule
        \textbf{Model} & \multicolumn{3}{c}{\textbf{Precision}} & \multicolumn{3}{c}{\textbf{Recall}} & \multicolumn{3}{c}{\textbf{Mean Average Precision}} \\
        \cmidrule(r){2-4} \cmidrule(r){5-7} \cmidrule(r){8-10}
        & P@10 & P@15 & P@20 & R@10 & R@15 & R@20 & MAP@10 & MAP@15 & MAP@20 \\
        \midrule
        $\text{CCA}_\text{SB+DW}$ & 0.1181 & 0.1008 & 0.09 & 0.1446 & 0.1819 & 0.2143 & 0.0789 & 0.0873 & 0.093 \\
        $\text{CCA}_\text{SB+N2}$ & 0.1185 & 0.1017 & 0.09 & 0.1472 & 0.1857 & 0.2164 & 0.0796 & 0.0883 & 0.0938 \\
        $\text{CCA}_\text{S2+DW}$ & 0.1843 & 0.1535 & 0.134 & 0.2301 & 0.2797 & 0.3195 & 0.1351 & 0.1484 & 0.1572 \\
        $\text{CCA}_\text{S2+N2}$* & 0.1859* & 0.1551* & 0.1348* & 0.2298* & 0.2799* & 0.3205* & 0.1348* & 0.1525* & 0.1611* \\
        $\text{DCCA}_\text{SB+DW}$ & 0.1211 & 0.1044 & 0.0924 & 0.152 & 0.191 & 0.2229 & 0.0819 & 0.0909 & 0.0965 \\
        $\text{DCCA}_\text{SB+N2}$ & 0.1218 & 0.1049 & 0.0928 & 0.152 & 0.1934 & 0.2251 & 0.0827 & 0.0918 & 0.0975 \\
        $\text{DCCA}_\text{S2+DW}$ & 0.1951 & 0.1629 & 0.1417 & 0.2473 & 0.2994 & 0.3406 & 0.1467 & 0.1612 & 0.1705 \\
        \textbf{$\text{DCCA}_\text{S2+N2}$} & \textbf{0.196} & \textbf{0.1639} & \textbf{0.1425} & \textbf{0.2478} & \textbf{0.3014} & \textbf{0.3422} & \textbf{0.1505} & \textbf{0.1653} & \textbf{0.1745} \\
        \midrule
        Rel. Gain (\%) & +5.43 & +5.67 & +5.71 & +7.83 & +7.68 & +6.77 & +11.64 & +8.39 & +8.32 \\
        \bottomrule
    \end{tabular}
    }
\end{table*}

Table \ref{tab:projection-concat-results} summarizes the performance of the fusion strategy that concatenates projected CCA and DCCA embeddings. The results demonstrate fusion based on CCA and DCCA consistently outperform simple concatenation, indicating that projecting embeddings into a shared subspace that maximizes correlation provides a more expressive fusion approach. Compared to single-view representations, CCA consistently improves performance across all metrics for both text and node models, except for MAP@10 with SPECTER2, where the text-only representation performs marginally better. These findings highlight the benefit of integrating complementary information from textual and structural views. Among fusion configurations, those that combine SPECTER2 text embeddings and Node2vec node embeddings yield the strongest results. This underscores the relative strength of Node2vec over DeepWalk in fusion contexts and reinforces the effectiveness of SPECTER2's citation-aware embeddings compared to SciBERT's general-purpose scientific language representations.

DCCA consistently outperforms CCA across all metrics and model combinations. Notably, when fusing SPECTER2 and Node2vec embeddings, DCCA achieves relative gains of over 11\% in MAP@10, 5\% in Precision@10, and 7\% in Recall@10. These results suggest that the relationship between textual and structural views involves complex, non-linear dependencies that linear CCA fails to capture. By leveraging neural networks to learn expressive non-linear transformations, DCCA effectively models these cross-modal interactions, leading to enhanced recommendation performance. However, the improvements offered by DCCA are less pronounced when applied to lower-quality base representations. The reduced gains observed with SciBERT-based configurations likely reflect the limitations of SciBERT's standalone embeddings, which provide weaker signals for cross-modal fusion and citation recommendation compared to SPECTER2-based embeddings. Since DCCA relies on the richness of its input views to learn effective joint representations, its ability to model complementary relationships is diminished when either modality carries limited or noisy information.

For comparison, we also evaluated a linear combination strategy on the DCCA embeddings, where fused representations were calculated as a weighted sum of the text and node projections using $\alpha \in \{0.1, 0.25, 0.5, 0.75, 0.9\}$. Although configurations favoring the text view, particularly $\alpha = 0.75$, achieved competitive performance, all variants consistently underperformed compared to projected concatenation. This indicates that linear weighting may obscure important cross-modal interactions and complementary signals that are better preserved through concatenation in a shared latent space. Full results are presented in Appendix \ref{sec:appendix-linear-combination}.

\section{Conclusion}
\label{sec:conclusion}
In this paper, we introduced a novel neural network-based algorithm that leverages DCCA to integrate textual and structural embeddings of scientific articles into a shared latent space. We investigated its application to the task of citation recommendation and demonstrated that it outperforms state-of-the-art CCA-based methods on a large-scale citation network dataset. Our results indicate that by capturing complex, non-linear relationships between complementary data views, DCCA facilitates more expressive cross-modal representations than linear CCA, leading to improved citation recommendation performance.

Our experiments showed that fusion strategies based on CCA and DCCA consistently outperform simple concatenation of unprojected embeddings and single-view representations. This highlights the benefit of learning a correlated subspace, which more effectively integrates textual and structural information. In particular, concatenating DCCA-projected embeddings yields stronger performance than linear combination strategies, likely because it better preserves cross-modal interactions. The advantages of DCCA are most evident when both input views are of high quality, for example, SPECTER2 for text and Node2vec for structure. However, its benefits are less pronounced with weaker base embeddings, such as those from SciBERT, underscoring the importance of strong representational inputs for successful multimodal fusion.

One limitation of our work is that network representation learning was restricted to homogeneous citation graphs, which constrained our ability to model the semantic and structural heterogeneity present in real-world academic networks. Future research could extend our approach by applying DCCA to integrate textual representations with node embeddings derived from heterogeneous graphs. Techniques such as metapath2vec \cite{Dong2017Metapath}, which utilizes meta-path-guided random walks, and HeGAN \cite{Hu2019HeGAN}, a graph adversarial learning approach, offer promising directions to model such graph structures. Additionally, while we demonstrated the benefits of concatenating DCCA-projected embeddings, alternative fusion strategies remain worth investigating. For example, utilizing an additional neural network to fuse the projections into a unified representation could further enhance performance by capturing deeper cross-modal relationships.

As the volume of scientific literature continues to grow, models that effectively combine textual and structural information will become increasingly essential for developing accurate and context-aware citation recommendation systems. Our work represents a step forward in this direction and opens avenues for future advancements in multi-view representation learning.

\bibliographystyle{unsrt}  
\bibliography{references}  

\appendix
\section{Linear Combination Fusion}
\label{sec:appendix-linear-combination}
\begin{table}[H]
    \caption{Feature fusion results using a linear combination of projected DCCA embeddings. S2 denotes SPECTER2, while DW and N2 refer to DeepWalk and Node2vec, respectively}
    \label{tab:linear-combination-results}
    \centering
    \resizebox{\textwidth}{!}{
    \begin{tabular}{llccc ccc ccc}
        \toprule
        \textbf{$\alpha$} & \textbf{Model} & \multicolumn{3}{c}{\textbf{Precision}} & \multicolumn{3}{c}{\textbf{Recall}} & \multicolumn{3}{c}{\textbf{Mean Average Precision}} \\
        \cmidrule(r){3-5} \cmidrule(r){6-8} \cmidrule(r){9-11}
        & & P@10 & P@15 & P@20 & R@10 & R@15 & R@20 & MAP@10 & MAP@15 & MAP@20 \\
        \midrule
        0.1 & $\text{DCCA}_\text{S2+DW}$ & 0.1598 & 0.1348 & 0.1177 & 0.1952 & 0.241 & 0.2757 & 0.1103 & 0.1221 & 0.1293 \\
        0.1 & $\text{DCCA}_\text{S2+N2}$ & 0.1624 & 0.1358 & 0.118 & 0.1978 & 0.2424 & 0.2763 & 0.1129 & 0.1243 & 0.1314 \\
        0.25 & $\text{DCCA}_\text{S2+DW}$ & 0.1684 & 0.141 & 0.1237 & 0.2068 & 0.2532 & 0.2906 & 0.1181 & 0.1303 & 0.1383 \\
        0.25 & $\text{DCCA}_\text{S2+N2}$ & 0.1705 & 0.1415 & 0.1235 & 0.2102 & 0.2541 & 0.2893 & 0.1211 & 0.1329 & 0.1406 \\
        0.5 & $\text{DCCA}_\text{S2+DW}$ & 0.1872 & 0.1558 & 0.136 & 0.2337 & 0.2866 & 0.3275 & 0.1374 & 0.151 & 0.1599 \\
        0.5 & $\text{DCCA}_\text{S2+N2}$ & 0.1875 & 0.156 & 0.1357 & 0.2354 & 0.2852 & 0.324 & 0.139 & 0.1524 & 0.1609 \\
        0.75 & $\text{DCCA}_\text{S2+DW}$ & 0.1906 & 0.1585 & 0.1376 & \textbf{0.2456} & 0.2986 & \textbf{0.34} & 0.144 & 0.1577 & 0.1659 \\
        0.75 & $\text{DCCA}_\text{S2+N2}$ & \textbf{0.1916} & \textbf{0.1596} & \textbf{0.1382} & 0.2453 & \textbf{0.2988} & 0.3394 & \textbf{0.1463} & \textbf{0.1599} & \textbf{0.1683} \\
        0.9 & $\text{DCCA}_\text{S2+DW}$ & 0.1833 & 0.1527 & 0.1329 & 0.2375 & 0.2896 & 0.3289 & 0.1407 & 0.1532 & 0.1611 \\
        0.9 & $\text{DCCA}_\text{S2+N2}$ & 0.1845 & 0.1536 & 0.1329 & 0.2385 & 0.2898 & 0.3285 & 0.1418 & 0.1546 & 0.1623 \\
        \bottomrule
    \end{tabular}
    }
\end{table}

Table \ref{tab:linear-combination-results} presents the results of the fusion strategy that linearly combines projected DCCA embeddings, using weighting coefficients $\alpha \in {0.1, 0.25, 0.5, 0.75, 0.9}$ to balance the contributions of the textual and structural views. We evaluated combinations of SPECTER2 text embeddings with DeepWalk and Node2vec node embeddings. Although configurations that emphasize the text view, especially $\alpha=0.75$, achieved competitive performance, linear combinations consistently underperform compared to concatenating the DCCA-projected embeddings, as shown in Table \ref{tab:projection-concat-results}. These findings suggest that linear weighting may result in a loss of complementary information that projected concatenation preserves more effectively.

\end{document}